\documentclass[pdftex,twocolumn,epjc3]{svjour3}          

\RequirePackage[T1]{fontenc}

\smartqed  

\RequirePackage{graphicx}
\RequirePackage{mathptmx}      
\RequirePackage{flushend}
\RequirePackage[numbers,sort&compress]{natbib}
\RequirePackage[colorlinks,citecolor=blue,urlcolor=blue,linkcolor=blue]{hyperref}

\journalname{}

\def\Journal#1#2#3#4{{#1} {\bf #2}, #3 (#4)}

\def\NIMA{{Nucl. Instrum. Methods} A}

\def\PLB{{Phys. Lett.}  B}

\def\PRD{{Phys. Rev.} D}

\def\PAN{Phys. At. Nucl.}

\def\JINST{JINST}

\def\JHEP{JHEP}

\def\EPJC{Eur. Phys. J. C}
\def\AstrPhy{Astropart. Phys.}

\def\aprle{\buildrel < \over {_{\sim}}}

\begin{document}

\title{Measurement of TeV atmospheric muon charge ratio  \\ 
with the full OPERA data}


\author{
N.~Agafonova\thanksref{MOSCOWINR}\and  
A.~Anokhina\thanksref{MOSCOWSINP}\and  
S.~Aoki\thanksref{KOBE}\and
A.~Ariga\thanksref{BERN}\and
T.~Ariga\thanksref{BERN}\and
D.~Bender\thanksref{ANKARA}\and 
A.~Bertolin\thanksref{PADOVAINFN}\and 
C.~Bozza\thanksref{SALERNO}\and  
R.~Brugnera\thanksref{PADOVA, PADOVAINFN}\and 
A.~Buonaura\thanksref{NAPOLI, NAPOLIINFN}
% ...
}

\thankstext[$\star$]{e1}{Corresponding authors. E-mail: mauri@bo.infn.it, sioli@bo.infn.it}
\thankstext{e3}{Now at Kyungpook National University, Daegu, Korea.}
\thankstext{e4}{Now at Samsung Changwon Hospital, SKKU, Changwon, Korea.}
\institute{
INR - Institute for Nuclear Research of the Russian Academy of Sciences, RUS-117312 Moscow, Russia\label{MOSCOWINR} \and
SINP MSU - Skobeltsyn Institute of Nuclear Physics, Lomonosov Moscow State University, RUS-119991 Moscow, Russia\label{MOSCOWSINP} \and
Kobe University, J-657-8501 Kobe, Japan\label{KOBE} \and
Albert Einstein Center for Fundamental Physics, Laboratory for High Energy Physics (LHEP), University of Bern, CH-3012 Bern, Switzerland\label{BERN} \and
METU - Middle East Technical University, TR-06531 Ankara, Turkey\label{ANKARA} \and
INFN Sezione di Padova, I-35131 Padova, Italy\label{PADOVAINFN} \and
Dipartimento di Fisica dell'Universit\`a di Salerno and ``Gruppo Collegato'' INFN, I-84084 Fisciano (Salerno), Italy\label{SALERNO} \and
Dipartimento di Fisica dell'Universit\`a di Padova, I-35131 Padova, Italy\label{PADOVA} \and
Dipartimento di Fisica dell'Universit\`a Federico II di Napoli, I-80125 Napoli, Italy\label{NAPOLI} \and
INFN Sezione di Napoli, 80125 Napoli, Italy\label{NAPOLIINFN}
% ...
}

\date{}


\maketitle

\emph{To the memory of Prof. G. Giacomelli}
\\

\begin{abstract}
The OPERA detector, designed to search for 
$\nu_{\mu} \to \nu_{\tau}$ oscillations in the CNGS beam, 
is located in the underground Gran Sasso laboratory, 
a privileged location to study TeV-scale cosmic rays.
For the analysis here presented, the detector  
was used to measure the atmospheric muon charge ratio in the TeV region. 
OPERA collected charge-separated cosmic ray data 
between 2008 and 2012. 
More than 3 million atmospheric muon events were detected and reconstruct-ed, among which about 110000 multiple muon bundles. 
The charge ratio $R_{\mu} \equiv N_{\mu^+}/N_{\mu^-}$ was measured separately for single and for multiple muon events. 
The analysis 
exploited the inversion of the magnet polarity which was performed on purpose during the 2012 Run. 
The combination of the two data sets 
with opposite magnet polarities 
allowed minimizing 
systematic uncertainties 
and 
reaching 
an 
accurate determination of the 
muon charge ratio. 
Data were fitted to obtain relevant parameters on the composition of primary cosmic rays  
and the associated kaon production in the forward fragmentation region. 
In the surface energy range 1-20 TeV investigated by 
OPERA,
$R_{\mu}$ is well described by a parametric model including only pion and kaon contributions to the muon flux, 
showing no significant contribution of the prompt component. The energy independence supports the validity of Feynman scaling 
in the fragmentation region up to $200$~TeV/nucleon primary energy. 

\end{abstract}

\section{Introduction}

Underground experiments 
detect the penetrating remnants of primary cosmic ray interactions in the atmosphere, 
namely 
muons and neutrinos.
These are the decay products of charged mesons contained in the particle cascade, mainly pions and kaons. 
At very high energies 
also 
charmed particles 
are expected to contribute. 

The muon charge ratio $R_{\mu} \equiv N_{\mu^+}/N_{\mu^-}$, defined as the number of positive over negative 
charged muons, is 
studied 
since many decades. 
It 
provides 
an understanding of the mechanism of multiparticle production in the atmosphere in kinematic regions not accessible to accelerators, as well as information on the primary cosmic ray composition. 
A charge ratio 
larger than unity 
reflects the abundance of protons over heavier nuclei  
in the primary cosmic radiation. 
The charge asymmetry is preserved in the secondary hadron production, and consequently in the muon fluxes, 
due to the steepness of the primary spectrum which enhances the forward fragmentation region~\cite{frazer}. 
The kaon contribution to the muon flux increases with the muon energy. 
Since the production of positive kaons is favoured by the associated production $\Lambda K^+$, the muon charge ratio is expected to rise with energy. 
Assuming the 
hypothesis of complete scaling 
we expect 
an energy independent charge ratio above the TeV energy region 
at sea level
\cite{frazer} once the kaon contribution to the muon flux  
reached its asymptotic value \cite{gaisser2}. 
At higher energies, 
around $O(100)$~TeV, 
the heavy flavor contribution, as well as changes in the primary composition, 
may 
become significant.

The OPERA experiment, described in detail in Ref.~\cite{OPERAdet}, 
is a hybrid electronic detector/emulsion apparatus, located in the underground Gran Sasso laboratory, at an average depth of 3800 meters of water equivalent. The main physics goal of the experiment is the first observation of neutrino oscillations in direct appearance mode in the $\nu_{\mu} \to \nu_{\tau}$ channel~\cite{tau1, tau2, tau3}. 
OPERA already reported a first measurement of the atmospheric muon charge ratio 
at TeV surface energies 
using the 2008 Run data~\cite{cosmic}. 
Here we present the final results obtained 
with the complete statistics.  
OPERA continuously accumulated cosmic ray data with the electronic detectors of the target 
over  
the whole year 
from 2008 up to 2012. 
However the magnetic spectrometers were 
active 
only 
during the CNGS Physics Runs, being  
switched off during the CNGS winter shutdowns. 

As it was done 
in Ref.~\cite{cosmic}, 
we used the momentum and charge reconstruction 
obtained via 
the Precision Trackers (PT) of the OPERA spectrometers~\cite{PT}. 
Layers of vertical drift tubes are arranged in PT stations instrumenting 
the two identical dipole magnets. 
The momentum and charge information  
is given by 
the angle $\Delta \phi$ in the bending plane, 
i.e. the difference between the track directions  
reconstructed by the two PT stations before and after each magnet arm. 
For nearly horizontal muons 
up to four 
bending angles 
can be measured 
in the two dipole magnets. 

\section{Data Analysis}
The cosmic ray data used for this analysis were  
collected during the five CNGS Physics Runs between 2008 and 2012. 
In the first four years (2008-2011) the magnetic field was  
directed upward 
in the first arm of both dipoles 
and in the opposite direction in the second arm (standard polarity, SP). 
In 2012 
the coil currents were reversed and the spectrometer operated 
in 
inverted polarity (IP) mode. 

A pre-selection was applied in order to select only 
stable conditions of detector operation. 
Short periods with increased electronic noise or with any subdetector under test were removed, as well as periods in which the magnets were not in nominal conditions. 
Details on atmospheric muon event selection, reconstruction and analysis can be found in Refs. \cite{cosmic, PhD}. 

The final SP data correspond 
to 625.0 live days, distributed among the Runs as shown in Table \ref{tab:exposure}. The final IP exposure 
is equivalent to  
234.8 live days. 

\begin{table}
\centering
\begin{tabular*}{\columnwidth}{@{\extracolsep{\fill}}cccc@{}}
\hline
Physics Run & Events & Bundles & Exposure (days) \\
\hline
 2008 & 403038 & 14576 & 113.3  (SP) \\ 
 2009 & 434214 & 17138 & 121.1  (SP) \\ 
 2010 & 616805 & 22427 & 172.6  (SP) \\ 
 2011 & 766554 & 28545 & 218.0  (SP) \\ 
 2012 & 823670 & 30976 & 234.8  (IP) \\ 
\hline
\end{tabular*}
\caption{Data sets with magnet configuration in standard polarity (SP) and inverted polarity (IP). For each Run 
the number of cosmic muon events, the number of muon bundles therein and the live time after the pre-selection of good quality running periods are reported. } \label{tab:exposure}
\end{table}

In the total SP + IP live time, 
3044281 cosmic muon events were recorded. 
Among them, 113662 are muon bundles, i.e. events with 
a 
muon multiplicity $n_{\mu}$ greater than 1. 
To reconstruct the muon charge, the track has to cross at least one magnet arm yielding a measurement of the bending angle $\Delta \phi$ by the PT system. 
This resulted in the reconstruction of momentum and charge for 
650492 muons in SP 
($28.7$~\% of the total muon events) 
and 244626 muons in IP
($28.9$~\% of the total muon events). 

In order to  
improve the charge identification purity, the two selection  
criteria used in Ref. \cite{cosmic} were applied to the data. 
The first selection is 
a track quality cut. 
The $\Delta \phi$ bending angle measurement is provided by the PT track reconstruction 
which is 
spoiled in events containing a large number of fired tubes, 
typically due to radiative processes.
When the number of PT hits is 
much larger than 
the number expected from geometrical considerations \cite{cosmic, PhD} 
the event is rejected. 

The second selection acts on the charge discrimination power. 
Events with a bending angle 
smaller than 3 
times the angular resolution 
were rejected. 
This corresponds to a maximum detectable momentum up to 1 TeV/c \cite{PhD}.  
A further  
cut was applied to remove a few events with 
very large deflections \mbox{($|\Delta \phi| > 100$~mrad)}, 
either due to the scattering of 
low momentum muons 
($p_{\mu} \leq 5$
~GeV/c)  
or 
mimicked by secondary particles produced in high energy events.

Muons induced by atmospheric neutrinos coming from below were removed from the data set 
on the basis of time-of-flight measurements. 
Contributions from muon backscattering or 
up-going charged particles induced by muons were computed 
according to Ref.~\cite{ambrosio} and found to be negligible.

The numbers of single and multiple muons surviving all the selection cuts and used in the computation of the muon charge ratio are reported in Table~\ref{tab:stat}. 

\begin{table}
\centering
\begin{tabular*}{\columnwidth}{@{\extracolsep{\fill}}ccccc@{}}
\hline 
\multicolumn{1}{c}{} &
\multicolumn{2}{c} 
{  
Single $\mu$} &
\multicolumn{2}{c}{  
Bundle $\mu$}\\
\hline
 & $N_{\mu^+}$ & $N_{\mu^-}$ & $N_{\mu^+}$ & $N_{\mu^-}$  \\
\hline
SP & 143628 & 105278 & 5252 & 4533 \\
IP & 53575  & 40086  & 1785 & 1740 \\
\hline
\end{tabular*}
\caption{Final statistics for the muon charge ratio measurement; the number of muons surviving the cuts is quoted for both 
magnet polarity configurations.  
For muon bundles we provide the total number of muons and not the number of events. }
\label{tab:stat}
\end{table}

\subsection{Systematic uncertainties and unbiased charge ratio}
\label{subsec:systematics}

The comparison between the two data sets with opposite magnet polarity (SP and IP) allows checking 
systematic uncertainties affecting the muon charge ratio. 
These can be 
cancelled out using a proper combination of the two data samples (see~\ref{app}). 

The two main sources of systematic uncertainties 
are due to alignment and charge misidentification. 

In principle a different acceptance for $\mu^{+}$ and $\mu^{-}$ could also contribute to the overall systematic uncertainty. 
However 
the symmetry of the detector geometry allows to safely neglect this contribution.  
An indirect confirmation is given by the compatibility of the charge ratio values computed separately in the two arms of the same magnet, 
where the magnetic field has opposite directions~\cite{PhD}.  

Using the SP and IP data sets, we checked the symmetry in the acceptance for each magnet arm. 
According to the reference frame defined in Ref.~\cite{cosmic}, where the $z$-axis points toward the CNGS direction, 
 muons travelling toward 
the positive $z$-axis are defined as \textit{south-oriented} (SO), while 
 muons travelling toward 
the negative $z$-axis are defined as \textit{north-oriented} (NO).  
A muon crosses 
a magnet arm in 
one of these two possible ``orientations''. 
South-oriented $\mu^{+}$ and north-oriented $\mu^{-}$ are deflected toward west in the first arm in SP mode. 
The reversals of 
either the muon incoming orientation 
or the polarity mode 
are equivalent ways to exchange the muon bending sign. 
We computed the ratio $A_i$ of the number $N_i$ of charge-reconstructed muons 
in SP mode to the number 
in IP mode (normalized by their 
polarity live time), $A_i = (N_i)_{SP}$/$(N_i)_{IP}$
for the two orientations in each magnet arm $i$. 
The results are reported in Table \ref{tab:acceptance}.
The values of $A_i$ obtained 
in one orientation 
are all compatible 
with the values obtained in the other orientation, as expected from a charge-symmetric spectrometer. 
The individual comparison between $A_i$(SO) and $A_i$(NO) 
for each arm 
disposes of 
possible small live time differences among PT stations. 
The results are consistent with unity within 
statistical errors. 

\begin{table*}
\centering
\begin{tabular*}{0.7\textwidth}{@{\extracolsep{\fill}}ccccc@{}}
\hline
 & $A_1$ & $A_2$ & $A_3$ & $A_4$ \\
\hline
SO & $ 1.009 \pm 0.010 $ & $ 0.991 \pm 0.009 $ & $ 0.997 \pm 0.009 $ & $ 1.005 \pm 0.009 $ \\
NO & $ 0.996 \pm 0.005 $ & $ 1.006 \pm 0.005 $ & $ 0.992 \pm 0.005 $ & $ 1.004 \pm 0.005 $ \\ 
\hline
\end{tabular*}
\caption{Ratio between SP and IP numbers of charge-reconstructed muons 
in each magnet arm. The normalization by the relative polarity live time is globally applied.} 
\label{tab:acceptance}
\end{table*}

We have 
investigated 
the systematic uncertainty related to the alignment of the PT system. 
The SP and IP bending angle distributions were compared separately for 
south- and north-oriented 
muons 
in each magnet arm.
In case of perfect alignment, the two distributions (normalized by their respective live times) would coincide. 
In the data, a systematic bending angle shift $|\delta \phi_s| \sim (0.10 \pm 0.03)$~mrad was observed 
on average (in each magnet arm, for $ i = 1, \dots ,4$:
$|\delta \phi_{s,i}| = \{<0.03, \, 0.07, \, 0.10, \, 0.15\}$ mrad). 
Inverting the muon orientation, $\delta \phi_{s,i}$ preserves the absolute value and flips the sign, 
as expected in case of misalignment. 
Note that the absolute value is compatible 
with the 
alignment systematic uncertainty $\delta \phi_{syst} = 0.08$~mrad given in \cite{cosmic}. 

The observed global shift 
$\delta \phi_s$ is however an average value. 
It is a cumulative result of local distortions, tilts and bendings which depend on the muon position, zenith and azimuth. 
We therefore did not globally correct for $\delta \phi_s$ 
since 
the combination of IP and SP data allows to 
completely remove this systematics at a local level. 
As detailed in the~\ref{app}, 
the unbiased charge ratio $\hat{R}_{\mu}$ is obtained by the normalized sum of $\mu^{+}$ over the normalized sum of $\mu^{-}$: 
\begin{equation} 
\hat{R}_{\mu} = \frac{\frac{N^+_{SP}}{l_{SP}} + \frac{N^+_{IP}}{l_{IP}}}{\frac{N^-_{SP}}{l_{SP}} + \frac{N^-_{IP}}{l_{IP}}} = \frac{R_{\mu} (1 - \eta) + \eta}{(1 - \eta) + \eta R_{\mu}}
\label{eq:combi}
\end{equation} 
where $l_{SP,IP}$ is 
the respective polarity live time 
and $\eta$ is the charge misidentification probability. 
This combination provides a charge ratio in which the effects induced by misalignments cancel out. 
Indeed, the last equation is exactly the relation between the reconstructed $\hat{R}_{\mu}$ and the true $R_{\mu}$  charge ratio in case of perfect alignment~\cite{cosmic}. 
Inverting this relation, the charge ratio $R_{\mu}$ is obtained from the measured $\hat{R}_{\mu}$ corrected by the misidentification probability. 

In principle, all the systematic contributions due to misalignment cancel with this combination of SP and IP data. 
The residual systematic errors which do not cancel 
are estimated by the 
difference between the charge ratio values computed separately for SO and NO orientations. 
Since the alignment bias has opposite sign in the two orientations, we take $|R_{\mu}(NO) - R_{\mu}(SO)|$ as the systematic uncertainty related to our combination procedure. 
It was found $\delta R_{\mu} = 0.001$ for single muon events and 
$\delta R_{\mu} = 0.013$ for multiple muon events. 
In the latter the statistical contribution is dominant. 

The second 
source of systematic uncertainty 
considered is related to the determination of 
$\eta$. 
The charge misidentification  
computed with Monte Carlo 
is $\eta_{\footnotesize{\textrm{MC}}} = 0.030$, 
nearly independent on the muon momentum in the range 
5 GeV/c $\aprle p_{\mu} \aprle$ 1 TeV/c~\cite{PhD}. 
We estimated the systematic uncertainty of $\eta$ using a subsample of experimental data, 
i.e. the muon tracks reconstructed in both arms of each spectrometer. 
The probability of wrong charge assignment was evaluated counting the fraction of tracks with different charges, 
and the experimental $\eta_{\footnotesize{\textrm{data}}}$ was derived. 
The difference between $\eta_{\footnotesize{\textrm{data}}}$ and $\eta_{\footnotesize{\textrm{MC}}}$ 
is $\delta \eta = 0.007 \pm 0.002$~\cite{cosmic}. 
This corresponds to a one-sided systematic uncertainty on the charge ratio $\delta R_{\mu} = 0.007$. 

The final systematic uncertainty is 
the quadratic sum of the misalignment and the misidentification contributions. 

\section{Results}

The charge ratio of single muons impinging on the 
apparatus was computed combining the two 
polarity data sets according to Eq.~\ref{eq:combi}. 
After the correction for charge misidentification and detector misalignment, 
the final measurement with the complete 5-year statistics yields the result: 
\[
R_{\mu} (n_{\mu} = 1) = 1.377 \pm 0.006 (stat.) ^{+0.007}_{-0.001} (syst.) 
\]
The charge ratio of multiple muon events was computed using all the muon charges reconstructed in events with $n_{\mu} > 1$. It is not computed within the bundle itself, but summing up all the positive 
and the negative charges belonging to the bundle subsample.
The result after polarity combination and correction for misidentification is 
significantly 
lower than the single muon value:
\[
R_{\mu} (n_{\mu} > 1)  = 1.098 \pm 0.023 (stat.)  ^{+0.015}_{-0.013} (syst.) 
\]
The smaller value of the charge ratio for multiple muon events originates 
from two effects. 
First, 
as pointed out in ~\cite{cosmic}, the multiple muon sample naturally selects heavier 
primaries, 
thus a neutron enriched primary beam 
($\langle A \rangle \simeq 3.4$ for single muons, $\langle A \rangle \simeq 8.5$ for bundles).  
Second, 
the selection of muon bundles 
biases the Feynman-$x$ distribution towards the central region ($x_F \simeq E_{\footnotesize{\textrm{secondary}}}/E_{\footnotesize{\textrm{primary}}} \rightarrow 0$), 
in which the sea quark contribution to secondary particle production becomes 
relevant~\cite{PhD}. 
Both processes cause a decrease in the charge ratio. 

The single muon charge ratio was projected at the Earth surface using a Monte Carlo based unfolding technique for the muon energy $\mathcal{E}_{\mu}$~\cite{PhD}. 
As a first attempt, 
only pion and kaon contributions to the total muon flux are considered. 
We used
the analytic approximation described in 
\cite{cosmic} 
to infer the 
fractions of charged mesons decaying into a positive muon, $f_{\pi^+}$ and $f_{K^+}$. 
This approach does not yet 
consider any energy dependence of the proton excess in the primary composition. 
In this case the  muon flux and charge ratio depend on the 
vertical surface energy $\mathcal{E}_{\mu} \cos \theta^*$, 
where $\theta^*$ is the zenith angle at the muon production point~\cite{lipari}. 

\begin{figure}
\begin{minipage}{\columnwidth}
\centering
\resizebox{\columnwidth}{!}{\includegraphics{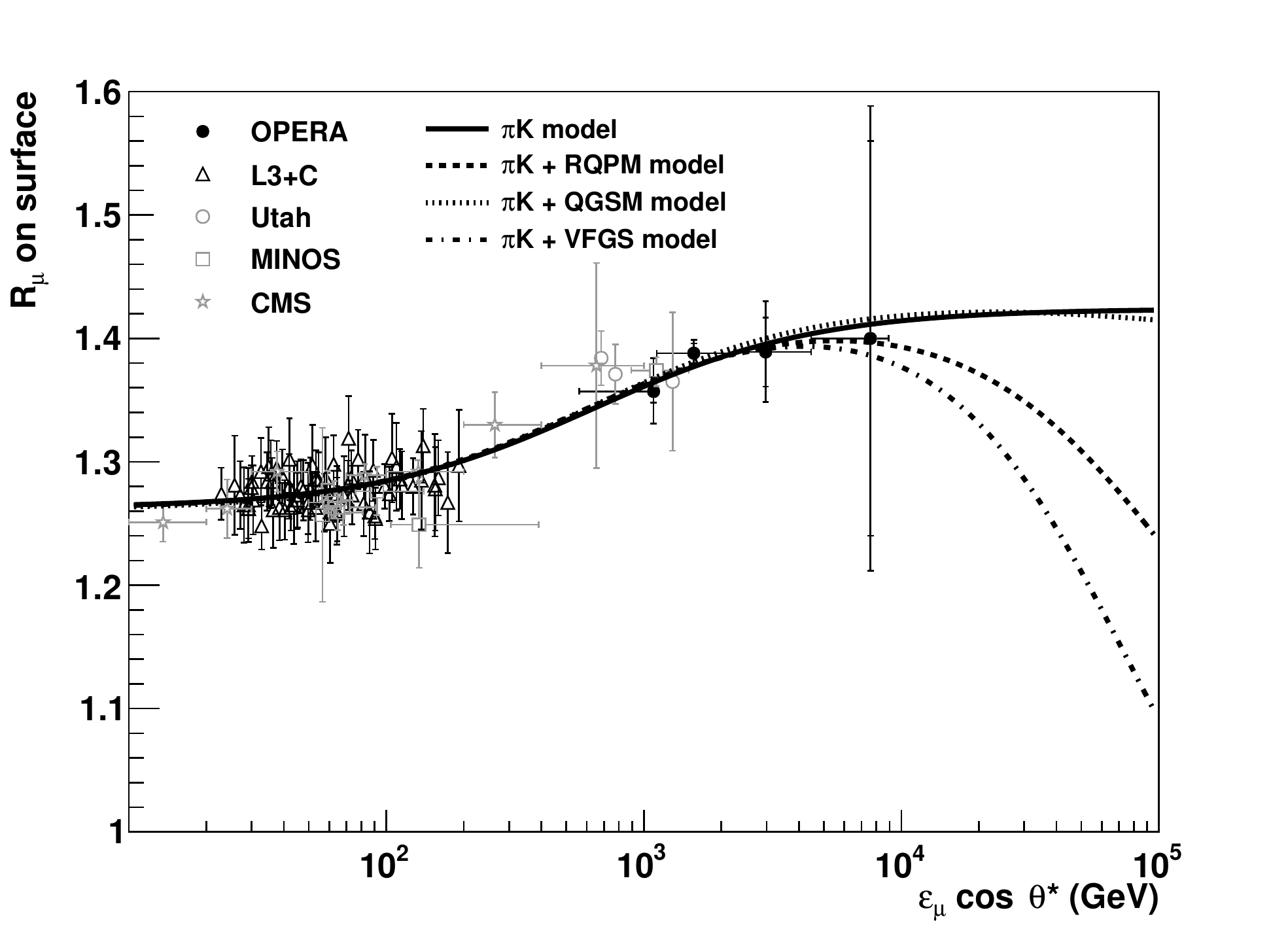} }
\end{minipage}
\caption{The muon charge ratio measured by OPERA as a function of the vertical surface energy $\mathcal{E}_{\mu} \cos \theta^*$ (black points).
Our data are fitted together with the L3+C~\cite{l3c} data (open triangles). 
The fit result is shown by the continuous line. The dashed, dotted and dash-dot 
lines are, respectively, the fit results with the inclusion of the RQPM~\cite{qgsm}, QGSM~\cite{qgsm} and VFGS~\cite{vfgs} models for prompt muon production in the atmosphere. The vertical inner bars denote the statistical uncertainty, the full bars show the total uncertainty. 
Results from other experiments, MINOS Near and Far Detectors~\cite{minos-nd, minos-fd}, CMS~\cite{cms} and Utah~\cite{utah}, are shown for comparison. }
\label{fig:emucos}
\end{figure}

$R_{\mu}$ is computed as a function of the vertical surface muon energy, binned according 
to the energy resolution, which is of the order of  d$(\log_{10} \mathcal{E}_{\mu}/\textrm{GeV}) \simeq 0.15$ in a logarithmic scale~\cite{PhD}. 
In each bin the two polarity data sets are combined and the obtained value is corrected for the charge misidentification. 
The two contributions to the systematic uncertainty are computed and added in quadrature. 
The results are shown in Fig.~\ref{fig:emucos}, together with data from other experiments (L3+C~\cite{l3c}, MINOS Near and Far Detectors~\cite{minos-nd, minos-fd}, CMS~\cite{cms} 
and Utah~\cite{utah}). 
The information for each of the four $\mathcal{E}_{\mu} \cos \theta^*$ bins are presented in Table \ref{tab:cr}: the energy range, the most probable value of the energy distribution in the bin, 
the average zenith angle, 
the charge ratio $R_{\mu}$, 
the statistical and systematic uncertainties. 

\begin{table*}
\centering
\begin{tabular*}{0.7\textwidth}{@{\extracolsep{\fill}}ccccccc@{}}
\hline 
\hline
Bin & $\mathcal{E}_{\mu} \cos \theta^*$ & $(\mathcal{E}_{\mu} \cos \theta^*)_{MPV}$ & $\langle \theta \rangle$ & $R_{\mu}$ & $\delta R_{\mu} (stat.)$ & $\delta R_{\mu} (syst.)$  \\
 & (GeV) & (GeV) & (deg) &  &  & \%  \\
\hline
1 & 562  - 1122 & 1091 & $ 47.5 $ & 1.357 & 0.009 & 1.8 \\
2 & 1122 - 2239 & 1563 & $ 42.8 $ & 1.388 & 0.008 & 0.1 \\
3 & 2239 - 4467 & 2972 & $ 46.9 $ & 1.389 & 0.028 & 2.1 \\
4 & 4467 - 8913 & 7586 & $ 60.0 $ & 1.40  & 0.16  & 7.1 \\
\hline
\hline
\end{tabular*}
\caption{The charge ratio in bins of $\mathcal{E}_{\mu} \cos \theta^*$. Here reported are the energy bin range, the most probable value of the energy distribution in the bin (MPV, evaluated using the full Monte Carlo simulation described in \cite{PhD}), the average zenith angle, 
the charge ratio and the statistical and systematic uncertainties. }
\label{tab:cr}
\end{table*}
Following the procedure described in \cite{cosmic}, we fitted our data and those from \cite{l3c} (for the high and low energy regions) 
in order to infer the fractions 
$f_{\pi^+}$ and $f_{K^+}$.  
In this approach, the atmospheric charged kaon/pion production ratio $R_{K/ \pi}$ had to be fixed. 
For this, we took the weighted average of experimental values reviewed in~\cite{seasonal}, $R_{K/ \pi} = 0.127$.  
The fit yields  
$f_{\pi^+} = 0.5512 \pm 0.0014$ and $f_{K^+} = 0.705 \pm 0.014$,  
corresponding to 
a muon charge ratio from pion decay $R_{\pi} = 1.2281 \pm 0.0007$ and a muon charge ratio from kaon decay $R_{K} = 2.39 \pm 0.07$.  

Taking into account various models for charm production, namely RQPM~\cite{qgsm}, QGSM~\cite{qgsm} and VFGS~\cite{vfgs}, 
the positive pion and kaon fractions 
obtained from the fit are unchanged within statistical errors. The results are shown in Fig.~\ref{fig:emucos}. 
The prompt muon component does not significantly contribute 
to $R_{\mu}$ 
up to \mbox{$\mathcal{E}_{\mu} \cos \theta^* \aprle 10$~TeV}. 

Recently, an enlightening analytic description 
of the 
muon charge ratio 
considering an explicit dependence on the relative proton excess in the primary cosmic rays, $\delta_0 = (p - n)/(p + n)$, 
was presented in \cite{gaisser2}: 
\begin{eqnarray}
R_{\mu} = \left[ \frac{f_{\pi^+}}{1 + B_{\pi} \mathcal{E}_{\mu} \cos \theta^*/\epsilon_{\pi}} + \frac{\frac{1}{2} (1+ \alpha_K \beta \delta_0)A_{K}/A_{\pi}}{1 + B^+_{K} \mathcal{E}_{\mu} \cos \theta^*/\epsilon_{K}} \right] \label{eq:fit} \\ \nonumber
\times \left[ \frac{1- f_{\pi^+}}{1 + B_{\pi} \mathcal{E}_{\mu} \cos \theta^*/\epsilon_{\pi}} + \frac{ (Z_{NK^-}/Z_{NK}) A_{K}/A_{\pi}}{1 + B_{K} \mathcal{E}_{\mu} \cos \theta^*/\epsilon_{K}} \right]^{-1} 
\end{eqnarray}
Here $p$ and $n$ fluxes are defined as 
\begin{equation}
p = \sum_i Z_i \, \Phi_i (E_N); \quad n = \sum_i (A_i - Z_i) \, \Phi_i (E_N)
\end{equation}
where the index $i$ runs over the primary ions (H, He, CNO, Mg-Si, Fe) and $E_N$ is the primary nucleon energy. 
The contributions from decays of pions and kaons 
are included in 
the kinematic factors $A_i, B_i, \epsilon_i \, (i = \pi, K)$ described in \cite{gaisser2, lipari}. 
An analogous contribution from charm decay is foreseen at high energies but still not observed. 
The spectrum weighted moments $Z_{ij}$~\cite{gaisser2} are 
contained in 
$\beta$ and $\alpha_K$: 
\begin{equation} 
\beta = \frac{1- Z_{pp} - Z_{pn}}{1- Z_{pp} + Z_{pn}}; \quad  \alpha_{K} = \frac{Z_{pK^+} - Z_{pK^-}}{Z_{pK^+} + Z_{pK^-}} 
\end{equation} 
Isospin symmetry allows expressing 
the pion contribution in terms of $f_{\pi^+}$, where
\begin{equation} 
f_{\pi^+} = \frac{1+\beta \delta_0 \alpha_{\pi}}{2}
\end{equation} 
Here $\alpha_{\pi}$ is obtained replacing the subscript $K$ with the subscript $\pi$ in $\alpha_K$. 

We extracted from 
the data the composition parameter $\delta_0$ and 
the factor $Z_{pK^+}$ related to the associated production $\Lambda \, K^+$ in the forward region. 
The $Z_{pK^+}$ moment is still poorly known 
and its predicted value considerably differs 
for different Monte Carlo codes~\cite{hadronic-models,lhc-cr}. 

In Eq.~\ref{eq:fit} the charge ratio does not exclusively depend 
on the vertical surface energy. 
Since the spectra of primary nuclei 
have different spectral indices, 
the parameter $\delta_0$ depends on the primary nucleon energy $E_N$. 
In the energy range of interest 
the approximation $E_N \simeq 10 \times \mathcal{E}_{\mu}$
can be used~\cite{gaisser2}. 

The correct way of taking into account the different dependencies is to simultaneously fit Eq.~\ref{eq:fit} as a function of the two variables $(\mathcal{E}_{\mu}, \cos \theta^*)$. 
In each $(\mathcal{E}_{\mu}, \cos \theta^*)$ bin the data sets with opposite polarities 
are combined and $\hat{R}_{\mu}$ is corrected for the charge misidentification. 

The pion moments $Z_{p \pi^+}$ and $Z_{p \pi^-}$ were set 
to 
the values reported in \cite{gaisser2}, 
since 
the fraction of positive pions in the atmosphere $f_{\pi^+} = 0.5512 \pm 0.0014$ 
derived in this work 
is robust and consistent with previous measurements~\cite{minos-nd, minos-fd} and with the $Z_{N\pi}$ values 
based on fixed target data~\cite{gaisser}. 
The moment $Z_{pK^-}$ was also set to 
the value given in \cite{gaisser2}, since for $K^-$ 
there is no counterpart of the 
associated production $\Lambda \, K^+$. 
On the other hand $K^-$ are equally produced in  $K^+ K^-$  pairs by protons and neutrons ($Z_{pK^-} \simeq Z_{nK^-}$). 

A linear energy dependence in logarithmic scale 
of the parameter $\delta_0$ was assumed, $\delta_0 = a + b \log_{10} (E_{N}$/GeV/nu-cleon$)$,
 as suggested by direct measurements of the primary composition and by the Polygonato model~\cite{polygonato}. 
We fixed the slope at 
$b = -0.035$ which was obtained fitting  
the values 
reported in \cite{gaisser2}. 
 
We made a two-dimensional fit of OPERA and L3+C data as a function of $(\mathcal{E}_{\mu}, \cos \theta^*)$ to Eq.~\ref{eq:fit} 
with $\delta_0$ and $Z_{pK^+}$ as free parameters. 
The fit 
yields the 
composition parameter at the average energy measured by OPERA $\langle \mathcal{E}_{\mu}  \rangle = 2$~TeV (corresponding to $\langle E_{N}  \rangle \approx 20$~TeV/nucleon) 
$\delta_0 (\langle \mathcal{E}_{\mu} \rangle ) = 0.61 \pm 0.02$ 
and the factor $Z_{p K^+} = 0.0086 \pm 0.0004$. 
 
\begin{figure}
\begin{minipage}{\columnwidth}
\centering
\resizebox{\columnwidth}{!}{\includegraphics{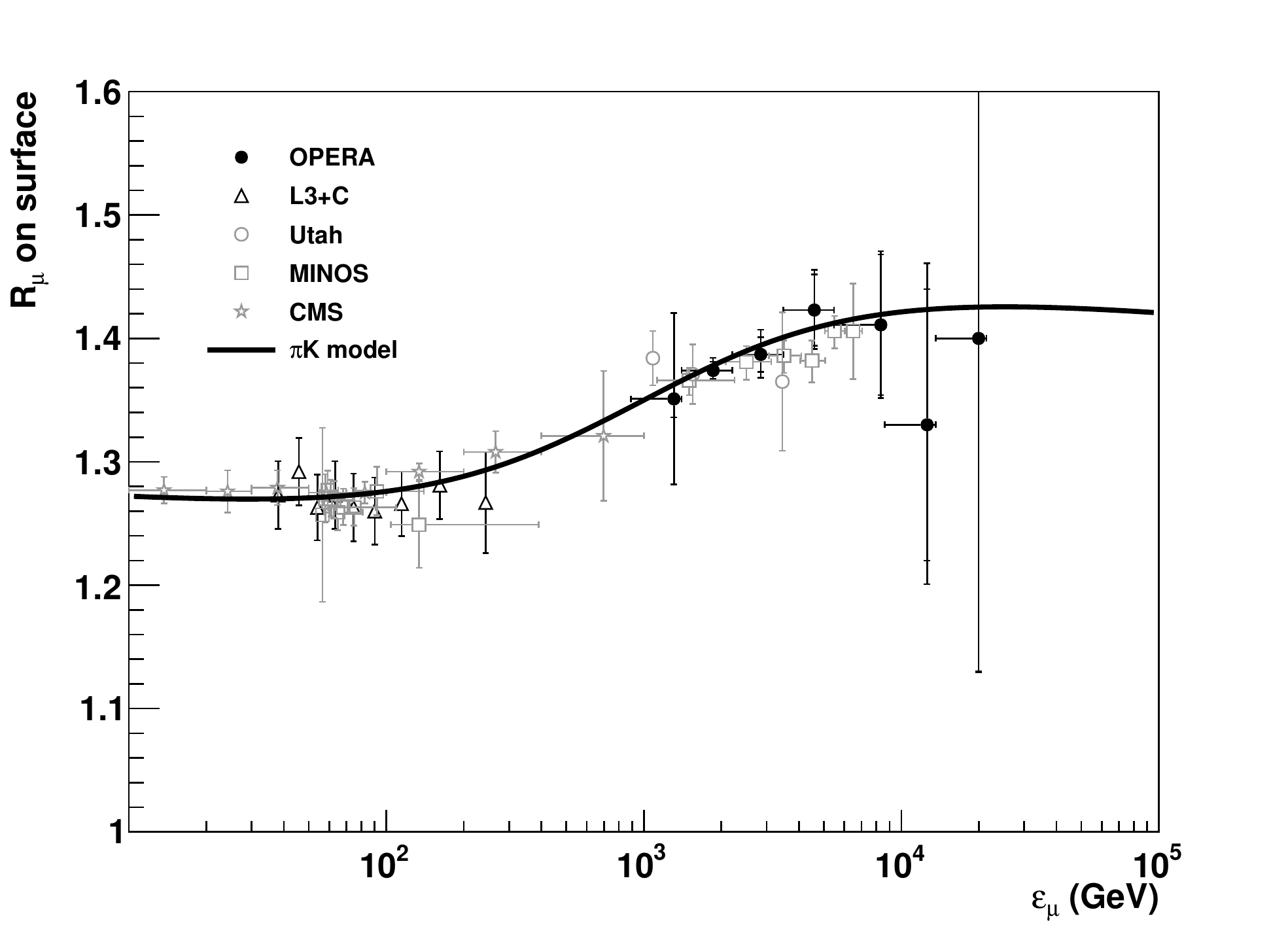} }
\end{minipage}
\caption{
Our measurement of the muon charge ratio as a function of the surface energy $\mathcal{E}_{\mu}$ (black points). 
The two-dimensional fit in $(\mathcal{E}_{\mu}, \cos \theta^*)$ yields a measurement of the composition parameter $\delta_0$ and of the factor $Z_{p K^+}$. 
The fit result is projected on the average OPERA zenith $\langle \cos \theta^* \rangle \simeq 0.7$ and shown by the continuous line. 
Results from other experiments, L3+C (only for $0.675 < \cos \theta < 0.75$)~\cite{l3c}, MINOS Near and Far Detectors~\cite{minos-nd, minos-fd}, CMS~\cite{cms} and Utah~\cite{utah},  
are also shown for comparison. }
\label{fig:emu}
\end{figure}

The result of the fit in two variables $(\mathcal{E}_{\mu}, \cos \theta^*)$ is projected on 
the average OPERA zenith $\langle \cos \theta^* \rangle \simeq 0.7$ 
and is shown in Fig.~\ref{fig:emu} together with the measured 
charge ratio as a function of the surface muon energy. 
The energy independence of the charge ratio above the 
TeV 
supports 
the validity of the Feynman scaling in the fragmentation region. 

\section{Conclusions}

The atmospheric muon charge ratio $R_{\mu}$ was measured 
with the complete statistics accumulated 
along the five years of data taking. 
The combination of 
the two data sets collected with 
opposite magnet polarities 
allows reaching 
the most accurate 
measurement in the high energy region to date. 
The underground charge ratio 
was evaluated 
separately for single and for multiple muon events. 
For single muons, the integrated $R_{\mu}$ value is 
\[
R_{\mu} (n_{\mu} = 1) = 1.377 \pm 0.006 (stat.) ^{+0.007}_{-0.001} (syst.)
\]
while for muon bundles  
\[
R_{\mu} (n_{\mu} > 1)  = 1.098 \pm 0.023 (stat.)  ^{+0.015}_{-0.013} (syst.)
\]
The integral value and 
the energy dependence of 
the charge ratio 
for single muons are compatible with 
the expectation from 
a simple model~\cite{gaisser2, gaisser} which takes into account only 
pion and kaon contributions to the atmospheric muon flux. 
We extracted the fractions of charged pions and kaons 
decaying into positive muons, 
$f_{\pi^+} = 0.5512 \pm 0.0014$ and 
$f_{K^+} = 0.705 \pm 0.014$.

Considering the composition dependence embedded in 
Eq.~\ref{eq:fit}, 
we inferred 
a proton excess in the primary cosmic rays $\delta_0 = 0.61 \pm 0.02$ at the energy $\langle E_{N} \rangle \approx 20$~TeV/nucleon 
and a spectrum weighted moment $Z_{p K^+} = 0.0086 \pm 0.0004$. 

The observed behaviour of $R_{\mu}$ as a function of the 
surface energy 
from 
$\sim 1$~TeV up to 20 TeV (about 200 TeV/nu-cleon for the primary particle) 
shows no deviations from 
a simple parametric model taking into account only pions and kaons as muon parents, 
supporting the hypothesis of limiting fragmentation 
up to primary energies/nucleon around 200 TeV.

\begin{acknowledgements}
We thank CERN for the successful operation of the CNGS facility and INFN for the 
continuous support given to the experiment during the construction, installation and commissioning phases through its LNGS laboratory. 
We warmly acknowledge funding from 
our national agencies: Fonds de la Recherche Scientifique-FNRS and Institut InterUniversitaire  
des Sciences Nucl\'eaires for Belgium, MoSES for Croatia, CNRS and IN2P3 
for France, BMBF for Germany, INFN for Italy, 
JSPS (Japan Society for the Promotion of Science), 
MEXT (Ministry of Education, Culture, Sports, Science and Technology), 
QFPU (Global COE programme of Nagoya University, Quest for Fundamental Principles 
in the Universe supported by JSPS and MEXT) and Promotion and Mutual Aid Corporation 
for Private Schools of Japan for Japan, 
SNF and the University of Bern 
for Switzerland, 
the Russian Foundation for Basic Research (grant no. 09-02-00300 a, 12-02-12142 ofim), 
the Programs of the Presidium of the Russian Academy of Sciences 
Neutrino physics and Experimental and theoretical research-es of fundamental interactions
connected with work on the accelerator of CERN, 
the Programs of Support of Leading Schools (grant no. 3517.2010.2), 
and the Ministry of Education and Science of the Russian Federation for Russia, 
the National Research Foundation of Korea Grant No. 2011-0029457 for Korea and TUBITAK, 
the Scientific and Technological Research Council of Turkey, for Turkey. 
We are also indebted to INFN for providing fellowships and grants to non-Italian researchers. 
We thank the IN2P3 Computing Centre (CC-IN2P3) for providing computing resources 
for the analysis and hosting the central database for the OPERA experiment.
We are indebted to our technical collaborators for the excellent quality of their work over 
many years of design, prototyping and construction of the detector and of its facilities.

\end{acknowledgements}

\appendix
\section{Combination of data sets}
\label{app}

A systematic shift of the bending angle distribution biases the integral value of the muon charge ratio. 
Moreover, since the unfolding of the surface muon energy is based on the underground muon momentum, 
a curvature bias 
has an important effect on the bin-to-bin migration matrix, 
i.e. the 
probability of measuring a surface energy $\mathcal{E}_{i}$ at a true energy $\mathcal{E}_{j}$. 

Due to misalignment 
there are 
in principle two different migration matrices $U^+$ and $U^-$ for each magnet polarity. 
Given the symmetry of the detector, 
the exchange of the magnet polarity is equivalent to the exchange of the charge sign (see Sect.~\ref{subsec:systematics}), 
thus $U^+_{SP} = U^-_{IP}$ and coherently $U^+_{IP} = U^-_{SP}$.

In general, with a curvature bias that shifts the bending angle distribution, a different charge misidentification $\eta^+$ and $\eta^-$ for positive and negative muons is 
expected 
for both standard and inverted magnet polarity. 
Given the symmetry of the detector, 
the relations $\eta^+_{SP} = \eta^-_{IP}$ and $\eta^-_{SP} = \eta^+_{IP}$ are valid. 
However we verified that after the application of the second selection criterion (the bending angle cut)
the charge misidentification $\eta$ 
is 
insensitive to the charge sign.  
We applied a rigid curvature bias $\delta \phi_s$ and observed that  
the bin construction 
clearly separates 
positive and 
negative bins. 
Therefore a symmetric misidentification $\eta = \eta^+ = \eta^-$ is assumed. 

Each energy bin content $N^{\pm}$ is the integral of the true charged muon flux $\Phi_{\mu}$ convolved with the migration matrix $U$ and corrected for the charge misidentification.  
For the standard polarity SP we have:
\begin{eqnarray} 
N^+_{SP} & = & \int_{E_1}^{E_2} dE^{'} \int_{-\infty}^{+\infty} [ U^+_{SP} (E,E^{'}) \Phi_{\mu}^+ (E) (1 - \eta) + \nonumber \\
&  & + U^-_{SP}(E,E^{'}) \Phi_{\mu}^- (E) \eta] \, dE  
\end{eqnarray} 
\begin{eqnarray} 
N^-_{SP} & = & \int_{E_1}^{E_2} dE^{'} \int_{-\infty}^{+\infty} [U^-_{SP} (E,E^{'}) \Phi_{\mu}^- (E) (1 - \eta) + \nonumber \\
&  & + U^+_{SP}(E,E^{'}) \Phi_{\mu}^+ (E) \eta]  \, dE  
\end{eqnarray} 
where $E_1, E_2$ are the lower and upper bounds of the reconstructed energy bin. 
The positive flux contribution can be rewritten in terms of the true charge ratio $R_{\mu}$ and the negative flux: 
\begin{equation} 
\Phi_{\mu}^+ (E) = R_{\mu} \, \Phi_{\mu}^- (E) 
\end{equation} 

Writing the same equations for the inverted polarity IP, 
the symmetries described above are taken into account:
\begin{eqnarray} 
N^+_{IP} & = & \int_{E_1}^{E_2} dE^{'} \int_{-\infty}^{+\infty} [ U^-_{SP} (E,E^{'})  R_{\mu} \, \Phi_{\mu}^- (E) (1 - \eta) + \nonumber \\
 & & + U^+_{SP}(E,E^{'}) \Phi_{\mu}^- (E) \eta]  \, dE 
\end{eqnarray} 
\begin{eqnarray} 
N^-_{IP} & = & \int_{E_1}^{E_2} dE^{'} \int_{-\infty}^{+\infty} [ U^+_{SP} (E,E^{'}) \Phi_{\mu}^- (E) (1 - \eta) + \nonumber \\ 
& & + U^-_{SP}(E,E^{'})  R_{\mu} \, \Phi_{\mu}^- (E) \eta]  \, dE 
\end{eqnarray} 

Thanks to 
the 
symmetric detector setup, 
the data combination able to cancel the misalignment systematic errors is the ratio 
$(N_{SP}^+ + N_{IP}^+)/(N_{SP}^- + N_{IP}^-)$,
where the numbers are normalized by the respective polarity live times.  
Indeed, writing the integrands only, we obtain: 
\begin{equation} 
N^+_{SP} + N^+_{IP} = (U^+_{SP}+ U^-_{SP}) (R_{\mu} \, \Phi_{\mu}^- (1 - \eta) + \Phi_{\mu}^- \eta )
\end{equation} 
\begin{equation} 
N^-_{SP} + N^-_{IP} = (U^-_{SP} + U^+_{SP}) (\Phi_{\mu}^- (1 - \eta) + R_{\mu} \, \Phi_{\mu}^- \eta ) 
\end{equation} 
Thus the unbiased charge ratio is given by the normalized sum of $\mu^{+}$ over the normalized sum of $\mu^{-}$: 
\begin{equation} 
\hat{R}_{\mu} = \frac{\frac{N^+_{SP}}{l_{SP}} + \frac{N^+_{IP}}{l_{IP}}}{\frac{N^-_{SP}}{l_{SP}} + \frac{N^-_{IP}}{l_{IP}}} = \frac{R_{\mu} (1 - \eta) + \eta}{(1 - \eta) + \eta R_{\mu}} 
\end{equation} 
The last equation is exactly the relation between the reconstructed $\hat{R}_{\mu}$ and true $R_{\mu}$  charge ratio in case of perfect alignment~\cite{cosmic}.


\begin{thebibliography}{99}

\bibitem{frazer} W.R. Frazer {\it et al.}, \Journal{\PRD}{5}{1653}{1972}.  
\bibitem{gaisser2} T.K. Gaisser, \Journal{\AstrPhy}{35}{801}{2012}.
\bibitem{OPERAdet} R. Acquafredda {\it et al.} (OPERA Collaboration), \Journal{\JINST}{4}{P04018}{2009}.
\bibitem{tau1} N. Agafonova  {\it et al.} (OPERA Collaboration), \Journal{\PLB}{691}{138}{2010}.
\bibitem{tau2} N. Agafonova  {\it et al.} (OPERA Collaboration), \Journal{\JHEP}{11}{036}{2013}.
\bibitem{tau3} N. Agafonova  {\it et al.} (OPERA Collaboration), arXiv:1401.2079, accepted by Phys. Rev. D. 
\bibitem{cosmic} N. Agafonova {\it et al.} (OPERA Collaboration), \Journal{\EPJC}{67}{25}{2010}. 
\bibitem{PT} R. Zimmermann {\it et al.}, \Journal{\NIMA}{555}{435}{2005}. 
\bibitem{PhD} N. Mauri, Ph.D. Thesis, Universit\`a di Bologna (2011), 
\\ http://operaweb.lngs.infn.it:2080/Opera/ptb/theses/theses/\\ Mauri-Nicoletta\_phdthesis.pdf
\bibitem{ambrosio} M. Ambrosio {\it et al.}, \Journal{\AstrPhy}{9}{105}{1998}. 
\bibitem{lipari} P. Lipari, \Journal{\AstrPhy}{1}{195}{1993}. 
\bibitem{hadronic-models} A. Fedynitch {\it et al.} \Journal{\PRD}{86}{114024}{2012}.
\bibitem{lhc-cr} A. Fedynitch, LHC-CR Workshop: Results and prospects of forward physics at the LHC (2013). 
\bibitem{gaisser} T.K. Gaisser, \textit{Cosmic Rays and Particle Physics}, Cambridge University Press (1990).
\bibitem{l3c} P. Achard {\it et al.} (L3+C Collaboration), \Journal{\PLB}{598}{15}{2004}.
\bibitem{minos-nd} P. Adamson {\it et al.} (MINOS Collaboration), \Journal{\PRD}{83}{032011}{2011}.
\bibitem{minos-fd} P. Adamson {\it et al.} (MINOS Collaboration), \Journal{\PRD}{76}{052003}{2007}.
\bibitem{cms} CMS Collaboration, \Journal{\PLB}{692}{83}{2010}. 
\bibitem{utah} G.K. Ashley II {\it et al.}, \Journal{\PRD}{12}{20}{1975}. 
\bibitem{seasonal} E.W. Grashorn {\it et al.}, \Journal{\AstrPhy}{33}{140}{2010}. 
\bibitem{qgsm} E.V. Bugaev {\it et al.}, \Journal{\PRD}{58}{054001}{1998}. 
\bibitem{vfgs} L.V. Volkova, G.T. Zatsepin, \Journal{\PAN}{71}{1782}{2008}. 
\bibitem{polygonato} J.R. H$\ddot{\textrm{o}}$randel, \Journal{\AstrPhy}{19}{193}{2003}. 

\end{thebibliography}
\end{document}